\documentclass[fleqn,usenatbib]{mnras}

\usepackage{newtxtext,newtxmath}

\usepackage[T1]{fontenc}

\DeclareRobustCommand{\VAN}[3]{#2}
\let\VANthebibliography\thebibliography
\def\thebibliography{\DeclareRobustCommand{\VAN}[3]{##3}\VANthebibliography}

\usepackage{graphicx}	
\usepackage{amsmath}	
\usepackage{siunitx}	
\usepackage{xcolor}		
\usepackage{physics}	
\usepackage{multicol}   
\usepackage{cleveref}   

\usepackage{ulem}
\newcommand{\old}[1]{}

\DeclareSIUnit[]{\mas}{\text{mas}}

\title[Different Approaches for SII]{Comparing Different Approaches for Stellar Intensity Interferometry}

\author[S. Karl et al.]{
Sebastian Karl,$^{1}$\thanks{E-mail: seb.karl@fau.de}
Andreas Zmija,$^{2}$\thanks{E-mail: andi.zmija@fau.de}
Stefan Richter,$^{1,3}$
Naomi Vogel,$^{2}$
Dmitry Malyshev,$^{2}$
\newauthor
Adrian Zink,$^{2}$
Thilo Michel,$^{2}$
Gisela Anton,$^{2}$
Joachim von Zanthier,$^{1,3}$
and Stefan Funk$^{2}$
\\
$^{1}$Institute for Optics, Information and Photonics, Friedrich-Alexander-Universität Erlangen-Nürnberg, Staudtstr. 1, Erlangen D 91058, Germany\\
$^{2}$Erlangen Centre for Astroparticle Physics, Friedrich-Alexander-Universität Erlangen-Nürnberg, Erwin-Rommel-Str. 1, Erlangen D 91058, Germany\\
$^{3}$Erlangen Graduate School in Advanced Optical Technologies, Paul-Gordan-Str. 6, Erlangen D 91052, Germany\\
}

\date{Accepted XXX. Received YYY; in original form ZZZ}

\pubyear{2021}

\begin{document}
\label{firstpage}
\pagerange{\pageref{firstpage}--\pageref{lastpage}}
\maketitle

\begin{abstract}
Stellar intensity interferometers correlate photons within their coherence time and could overcome the baseline limitations of existing amplitude interferometers. Intensity interferometers do not rely on phase coherence of the optical elements and thus function without high grade optics and light combining delay lines.
However, the coherence time of starlight observed with realistic optical filter bandwidths ($> \SI{0.1}{\nano \m}$) is usually much smaller than the time resolution of the detection system ($> \SI{10}{\pico \s}$), resulting in a greatly reduced correlation signal. Reaching high signal to noise in a reasonably short measurement time can be achieved in different ways: either by increasing the time resolution, which increases the correlation signal height, or by increasing the photon rate, which decreases statistical uncertainties of the measurement. 
We present laboratory measurements employing both approaches and directly compare them in terms of signal to noise ratio. A high time-resolution interferometry setup designed for small to intermediate size optical telescopes and thus lower photon rates (diameters $<\,$some meters) is compared to a setup capable of measuring high photon rates, which is planned to be installed at Cherenkov telescopes with dish diameters of $> \SI{10}{\m}$.
We use a Xenon lamp as a common light source simulating starlight. Both setups measure the expected correlation signal and work at the expected shot-noise limit of statistical uncertainties for measurement times between \SI{10}{\min} and \SI{23}{\hour}. We discuss the quantitative differences in the measurement results and give an overview of suitable operation regimes for each of the interferometer concepts.
\end{abstract}




\section{Introduction}
In recent years, interest in intensity interferometry for astronomical observations has grown. Prospects of precise measurements of angular diameters of stars and resolving structures on stellar surfaces \citep{Nunez2012, Nunez2015, Dravins2013} resulted in a revival of Hanbury-Brown Twiss interferometry in astrophysics.

High optical resolutions of less than \SI{1}{\mas} require interferometers with baselines of several hundred metres when measuring at optical wavelengths. Despite the problems due to atmospheric turbulences and the challenge of combining the light paths from the telescopes with sub-wavelength precision, stellar interferometers based on amplitude interferometry provide successful measurements at such baselines, see for example \cite{Roettenbacher2016, Che2011, Baines2007, Mourard2015}. The Center for High Angular Resolution Astronomy (CHARA) - one of the largest amplitude interferometers - has baselines of around \SI{300}{\m} providing angular resolution capabilities up to \SI{0.4}{mas} \citep{CHARA_description}.\\

The concept of intensity interferometry is promising for even longer baselines. Correlating the light intensities rather than amplitudes provides a robust way to overcome atmospheric disturbances at the cost of the achievable limiting magnitudes. Pioneering research at the Narrabri Stellar Intensity Interferometer was carried out by Hanbury Brown and Twiss (HBT) \citep{BROWN1956, brown1967stellar, HBT_32} in the second half of the 20th century. 
The challenge in this technique is the low amount of thermal photons that can actually interfere compared to the total number of arriving photons, increasing the difficulty in retrieving the signals in HBT interferometry.
For this kind of photon correlation measurement, high electronic stability, photo-detectors with high time resolution and large light collection areas (telescopes) are desirable. These are preconditions that are well achieved by today's telescopes and photo-detection equipment.
 
The performance of intensity interferometers is determined by a wide range of parameters, such as the aperture of the telescopes, the optical bandwidth of the correlated light, the timing resolution of the detection system, the maximum possible photon rate, and the projected baseline, all of which influence the expected signal-to-noise-ratio (S/N) of the measurement. However, when scanning the available parameter space for stellar intensity interferometry measurements at existing or future observatories, it appears that there are two promising methods for enabling successful measurements of the correlation signal.

The first method attempts to reach high timing resolution ($< \SI{1}{\nano \s}$) and narrow optical bandwidths ($< \SI{1}{\nano \m}$), keeping the correlation signal as pronounced as possible. This approach requires optical telescopes with photon timing dispersion smaller than the timing resolution of the detection system. Such telescopes typically enable precise control of the light path necessary for sharp optical filtering. As a downside, they usually do not come with large photon collection areas resulting in smaller photon count rates at the detectors compared to possible intensity interferometers at the world's largest telescopes. Also, high timing resolution photon counting detectors lose efficiency when operated at photon rates higher than some (tens of) MHz due to the detector dead time, reducing the detected photon rate further. This may lead to rather long measurement times of up to a few hours for measuring significant intensity correlations even with relatively bright stars of apparent magnitude 3. An option to overcome this downside is using arrays of telescopes, where each subset measures correlations along the same baseline, or spectral and polarization multiplexing \citep{Kim2014}. In absence of arrays, the main challenge is to obtain stable measurement conditions over periods of up to several hours.
In this context, some promising results have been achieved in recent years. Temporal photon bunching peaks of $\alpha$ Boo, $\alpha$ CMi and $\beta$ Gem, providing magnitudes between $-1.68$ and $-0.11$ in the near-infrared, were measured at the C2PU observatory in France with a telescope diameter of not more than $1\,$m at integration times between $2$ and $7$ hours \citep{guerin2017temporal}. Using two of these telescopes enabled spatial bunching measurements of $\alpha$ Lyr, $\beta$ Ori and $\alpha$ Aur \citep{guerin2018spatial}. The same telescopes were recently used to estimate the distance to P Cygni using intensity interferometry \citep{Rivet2020}. Further, the AQUEYE+ and IQUEYE astronomical high-timing resolution instruments can be used for stellar intensity interferometry at sub-nanosecond time resolution \citep{aqueye_ii}. Finally intensity interferometry observations and diameter estimations on 5 stars were performed by \cite{Horch2021} at the Southern Connecticut Stellar Interferometer.

In the second method, very large telescopes or telescope arrays are used. In this context, Imaging Air Cherenkov telescopes (IACTs) are promising candidates for photon correlation measurements \citep{Bohec2006}, since they provide light collection areas of about \SI{100}{\m ^2} (e.g. VERITAS \citep{maier2007veritas} or H.E.S.S. \citep{bernlohr2003optical}). Depending on the optical design Cherenkov telescopes have different levels of internal unisochronicity that make it impossible to achieve timing resolutions below these time differences.
While Davies-Cotton telescopes, such as VERITAS \citep{weekes2002veritas} and H.E.S.S \citep{bernlohr2003optical} incur time differences in the light path of a few nanoseconds, parabolic IACTs such as MAGIC \citep{shayduk2005new} or the Cherenkov Telescope Array (CTA) Large-Sized Telescope (LST) reduce it to $\approx 100-400\,$picoseconds, which is still significant compared to imaging-grade telescopes.
In addition, imperfect optics make it hard to collimate the light for optical filters, requiring bandwidths of \SI{1}{\nano \m} or more if interferometric filters are used. Photo-detection for such telescopes needs to be designed to work at several hundreds of MHz. Even though the correlation signal is very small in this arrangement the high photon rates suppress the shot noise enabling high S/N in relatively short measurements times of a few minutes. The main challenge here is the electronic stability on a high statistical level. Using the MAGIC telescopes \cite{magic_ii} measured photon bunching signals of a few stars within minutes (for single baselines). \cite{veritas_II} used the four VERITAS telescopes to measure the angular diameter of two stars by taking measurements at many different projected telescope baselines with averaged measurement times of \SI{17}{\min} per baseline, demonstrating the potential of IACTs for stellar intensity interferometry.

In this paper we shortly recapitulate the theory of intensity correlations in \cref{sec:2}. In \cref{sec:3} we describe the setup used to compare our complementary approaches in the laboratory. We then evaluate our measurement results in \cref{sec:4}. Finally, we compare and discuss our measurement results in \cref{sec:5}, illustrating the scope of both methods. Here we pay special attention to the S/N of our measurements, since it is the central parameter to quantify the quality of photon correlation measurements. High timing resolutions as well as high photon rates favour a high S/N, however, both are difficult to achieve simultaneously with current state-of-the-art technology.
We interpret our measurements in order to quantify telescope layouts for both approaches which are promising for high S/N. Further we will discuss future observation possibilities based on upcoming detector developments, which will especially boost the HTR technology towards higher photon rate capabilities and thus larger telescopes.

\section{Theory of intensity correlations}
\label{sec:2}
The observable of intensity interferometry is the normalized second order correlation function 

\begin{equation}
g^{(2)}(\mathbfit{r}, \tau) = \frac{\expval{I(\mathbfit{R}, t) I(\mathbfit{R} + \mathbfit{r}, t + \tau)}}{\expval{I(\mathbfit{R}, t)} \expval{I(\mathbfit{R} + \mathbfit{r}, t + \tau)}},
\end{equation}
where $I(\mathbfit{R}, t)$ denotes the intensity measured at position $\mathbfit{R}$ at a time $t$ and the brackets denote a time average. For thermal light sources (TLS) the correlations in space and time can be factorized into the respective absolute value squared of the first order correlation functions \citep{Mandel1995}:

\begin{equation}
    g^{(2)}(\mathbfit{r}, \tau) = 1 + \abs{g^{(1)}(\mathbfit{r})}^2 \abs{g^{(1)}(\tau)}^2.
    \label{eq:g2_factorization}
\end{equation}
Here $\abs{g^{(1)}(\mathbfit{r})}^2$ corresponds to the modulus squared of the measured visibility of amplitude interferometry \citep{Labeyrie2006} and is related to the source geometry by the Fourier relationship of the van Cittert Zernike theorem \citep{Mandel1995}. In the following, the correlation functions' spatial part is assumed to be $1$, as we restrict ourselves to correlations of two detectors having zero separation. In analogy to the spatial part, the temporal first order correlation can be calculated using a Fourier transform  relationship - the Wiener Khintchine theorem \citep{Mandel1995}:

\begin{equation}
    g^{(1)}(\tau) = \int_{-\infty}^\infty s(\omega) \mathrm{e}^{-i \omega \tau} \dd{\omega}
\end{equation}
where $s(\omega)$ denotes the normalized source spectrum. 

In general, TLS show bunching, meaning it is more likely to detect two photons arriving coincidentally rather than individually for a sufficiently small coincidence window. This is signified by the second order correlation function having its' maximum $g^{(2)}(\tau = 0) = 2$ at zero time delay. For a time delay much larger than the source coherence time $\tau_c$, the second order correlation function drops off to its baseline value of 1. The corresponding time delay is proportional to the source coherence time $\tau_c$, which can be defined as the integral over the correlation peak if the light is considered to be fully polarized

\begin{equation}
    \tau_c = \int_{-\infty}^\infty \abs{g^{(1)}(\tau)}^2  \dd{\tau} \approx \frac{\lambda^2}{c \Delta \lambda}. 
    \label{eq:coh_time_def}
\end{equation}
According to \cref{eq:coh_time_def} the coherence time is approximated in terms of the spectrum's central wavelength $\lambda$, the spectral bandwidth $\Delta \lambda$ and the speed of light $c$ \cref{eq:coh_time_def} \citep{Fox2006}. For a very narrow rectangular filter, the approximate equality in  \cref{eq:coh_time_def} holds rigorously.

For a real-world correlation measurement, the theoretical correlation function from \cref{eq:g2_factorization} has to be convolved with the detector response to yield the measurement expectation. Thus, for an electronic time resolution $\tau_\textnormal{e}$, the peak amplitude $A$ is reduced to a value $A = \frac{\tau_\textnormal{c}}{\tau_\textnormal{e}}$ from its initial value of 1 \citep{Loudon2000}. A further decrease by a factor of $\frac 1 2 $ is incurred if no polarization filtering is employed, as will be the case in our measurements. 

Assuming only statistical fluctuations the S/N of the HBT bunching peak of randomly polarized thermal light is given by \citep{Brown1957}

\begin{equation}
    S/N = A  \,  n(\omega) \, \eta  \, \sqrt{\frac{T_{\mathrm{meas}}}{\tau_\textnormal{e}}}, 
    \label{eq:SNR_def}
\end{equation}
where $A$ denotes the telescope collection area, $n(\omega)$ denotes the source spectral flux density, $\eta$ denotes the detector quantum efficiency, and $T_{\mathrm{meas}}$ denotes the measurement time. Note that the spatial correlation factor is 1 in our measurements and thus is omitted in further calculations. \Cref{eq:SNR_def} shows that for a HBT experiment, the filter bandwidth does not influence the S/N measurement. It should however be noted that \cref{eq:SNR_def} is only valid if the measured photon rate is approximately constant over the whole measurement time. This is not the case for both the single photon and current correlation measurements that are presented in this paper.
Instead we estimate the theoretical S/N utilizing shot noise as a lower limit to the noise that might be incurred in the noisy correlation measurement.

\section{Methods}
\label{sec:3}
Our two systems will be denoted as
\begin{itemize}
    \item HTR : "High time resolution" - photon counting measurements aiming for very high time resolutions, but restricted in photon rate capability.
    \item HPR : "High photon rates" - Photomultiplier tube (PMT) current measurements aiming for very high photon rates, but restricted in time resolution.
\end{itemize}

\begin{figure}
    \includegraphics[width=\columnwidth]{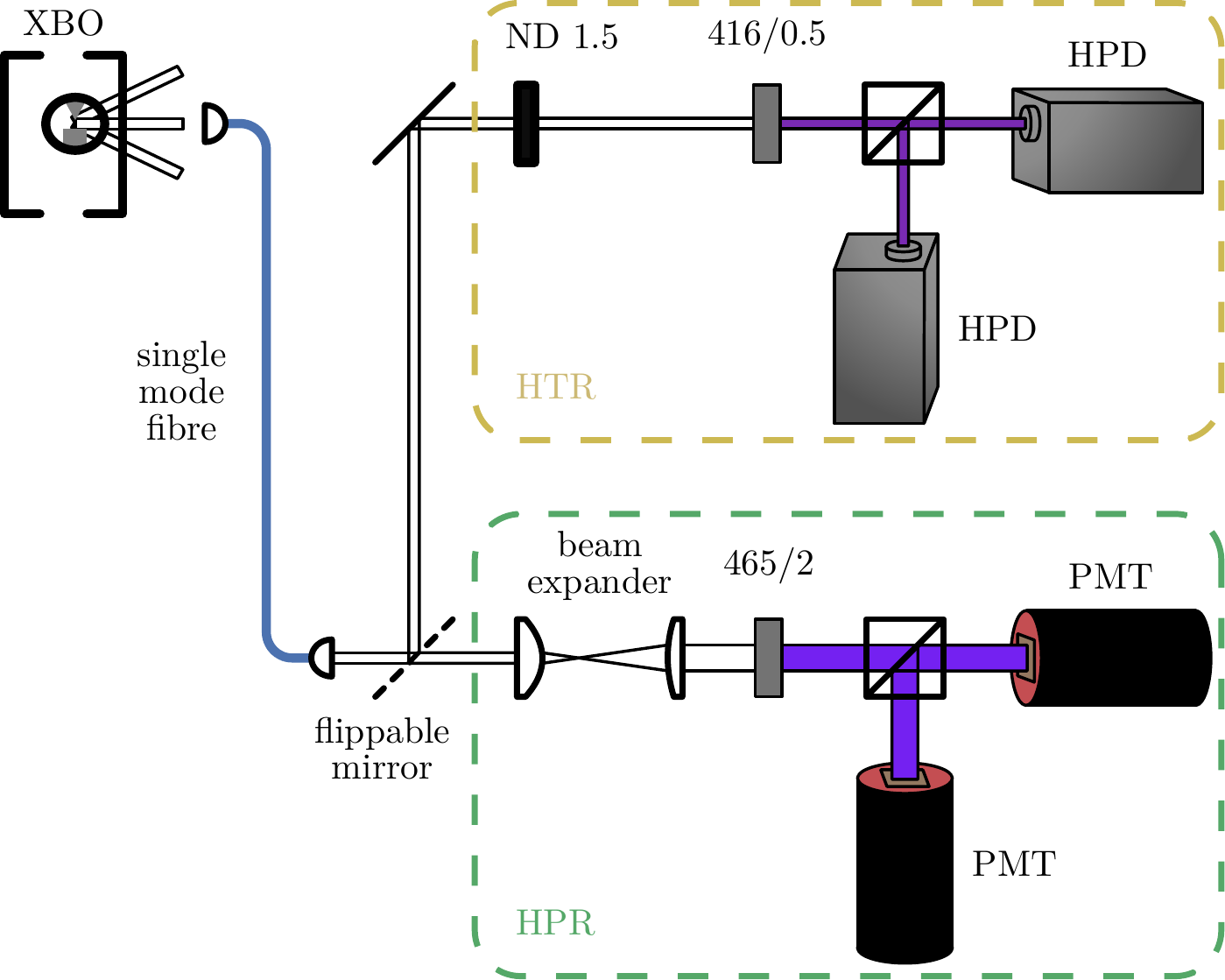}
    \caption{Measurement setup including both the HTR and the HPR parts, which can be swapped in operation by either installing or removing the flippable \SI{45}{\degree} mirror. The interference filters are denoted as follows: central wavelength/FWHM bandwidth. The high pressure xenon arc lamp is denoted by XBO, and the Photomultiplyer tubes and the hybrid photo detectors are denoted as PMT and HPM respectively. ND 1.5 refers to a neutral density filter of optical density 1.5 placed in the HTR setup.}
    \label{fig:setup}
\end{figure}

\Cref{fig:setup} shows a sketch of the combined setup. To simulate light coming from a distant star (black body spectrum, spatially coherent), a high pressure xenon arc lamp (XBO) is used \citep{Tan2014}. Light from this black body spectrum source is coupled into a single mode fiber placed at an appropriate distance to the XBO, ensuring the collection of only one spatial mode. After the light is transported through the single mode fiber, it is collimated using an achromatic fiber port.
The collimated beam can now be transferred to any of the setups by installing the flippable \SI{45}{\degree} mirror into the beam for the HTR setup, and removing it from the beam for the HPR setup. Details of both setups are described in the following subsections.

\subsection{The HTR setup}
The HTR setup is designed for use on optical telescopes, imprinting an optical time spread of not more than \SI{100}{\pico \s} onto the incident photons, and yielding light at a low intensity of less than \SI{10}{\mega \hertz} when filtered down to bandwidths $\le \SI{1}{\nano \m}$. In this case, the second order correlation function can be measured using single photon counting detectors and calculated from the measured arrival time differences.

We use two HPM 100-06 Hybrid Photo detectors (HPDs) manufactured by Becker\&Hickl to measure the temporal correlation function using the HBT setup. The photon arrival signals are discretized and correlated on a SPC-130-EMN time-to-amplitude-converter (TAC) based card manufactured by Becker\& Hickl. As these detectors have a maximum continuous count rate of \SI{15}{\mega \hertz}, usage at higher count rates within this regime results in undetected photons due to their $\approx \SI{70}{\nano \s}$ dead time. We measured the full width at half maximum (FWHM) overall system timing resolution to be \SI{41.61 +- 0.08}{\pico \s} and the detectors quantum detection efficiency to be 21 percent in the relevant wavelength range, in close agreement to the manufacturer specifications \citep{BHG2017}. Considering that the detector dead time is orders of magnitude larger than the final histogram length, simple start stop histogram correlation is sufficient to calculate the correlation function accurately and thus TAC based correlation can be utilized. To shift the signal out of the TAC switching systematics and eliminate the loss of correlation events due to the asymmetry of the TAC input channels, a cable delay of \SI{35}{\nano s} is introduced.

Since the HPM detectors have a maximal photon rate of \SI{15}{\mega \hertz}, the light directed into the HTR setup is attenuated by a neutral density (ND) filter of optical density 1.5, resulting in a 97.2 percent decrease in incident photon flux \citep{ThorlabsNDFilter}. Afterwards the attenuated light is spectrally filtered using a narrow interference filter of central wavelength \SI{416}{\nano m} FWHM width \SI{0.5}{\nano \m} manufactured by Alluxa and directed onto the HPM cathodes using a non-polarizing 50:50 beam splitter. 

In addition to pronounced switching systematics, the TAC imprints further systematics on the correlation measurements, consisting of two superimposed oscillations and an increasing slope. Since these systematics cannot be neglected, the measurements have to be calibrated. The calibration measurement is performed using light from a halogen light bulb, which is not spectrally filtered and incoherently scattered into a multimode fibre connected to our detectors. The measurement produces a nearly flat correlation histogram, with the HBT peak suppressed by a factor $\approx \SI{5e-5}{}$ compared to the visibility measured with spectral and spatial filtering. Subtracting the calibration measurement from the actual measurement yields a well resolved correlation peak. This correlation peak is then shifted by $\approx \SI{35}{\nano \s}$ to account for the cable delay.

\subsection{The HPR setup}
The HPR setup is designed to be operated at Imaging Air Cherenkov Telescopes (IACTs) and therefore deals with different parameters. Filter bandwidths of not less then \SI{2}{\nano \m} and thus photon rates of more than \SI{100}{\mega \hertz} even for stars of intermediate magnitude are realistic.
Typical telescope timing precisions of a few ns due to the unisochronicity of different light paths in the telescopes decrease the measured correlation signals strongly, however, the high photon fluxes of several hundreds of MHz at the detectors decrease shot noise maintaining a reasonable S/N.

At photon rates of hundreds of MHz, time stamping of single photons becomes difficult. Instead, the photon current of the photo-detector is measured in both observation channels. The interferometer consists of two Photomultiplier Tubes of type Hamamatsu R11265U, whose output current is amplified by a factor of ten and then digitized with \SI{1.6}{\nano \s} sampling time using the M4i.2212-x8 digitizer card from Spectrum Instrumentation.

The incoming beam from the single mode fiber is expanded to a diameter of $\approx \SI{1}{\centi \m}$ using a Keplerian beam expander, in order to not damage the photo-cathodes due to the high light density of the former beam.

If $\mathbfit{A}(t)$ and $\mathbfit{B}(t)$ are the vectors of the time-sampled photo-currents with subtracted mean from the two PMT channels, then the cross correlation with time shift $\tau$ is determined as:

\begin{equation}
    G^{(2)} (\tau) = \mathbfit{A} (t) \cdot \mathbfit{B} (t+\tau)
\end{equation}

Normalization is done by dividing every value by the mean value of the correlation baseline, where the baseline is defined as $G^{(2)}(\tau)|_{\tau \gg \tau_C}$.

\begin{equation}
    g^{(2)} (\tau) = \frac{G^{(2)} (\tau)}{\overline{G^{(2)}}|_{\tau \gg \tau_c}}
\end{equation}

This correlation procedure and the fact, that single photon pulse shapes are extended over $>30$ sampling time bins induces correlation effects in the $g^{(2)}$ of neighbouring bins. The baseline in part a) of \cref{fig:ECAP} shows that neighbouring $g^{(2)}$ values are not independent of each other. This correlation also results in a decreased value of the root mean square error (RMSE) expectation level compared to Poissonian shot noise calculations. A correction factor is obtained by averaging over experimental RMSE observations at short measurement times and confirmed by waveform simulations. In particular, the entire \SI{10}{\min} of measurement time are divided in $175$ chunks each of $\approx \SI{3.4}{\s}\,$ length. At these short time scales the RMSE is assumed to be purely statistical for $\tau \gg \tau_\textnormal{c}$. We further cross check the calculation of the correction factor by waveform simulations which then are correlated.
These simulations consist of an "empty" waveform baseline to which photons are added via a photon pulse shape template created by averaging many single photon pulse shapes measured at low rates. The simulated photon heights also follow measured photon pulse height distributions at low rates where the heights of single photons can be determined. No photon bunching is simulated in the waveforms, but the RMSE of the resulting correlation can be compared to the expectation of the single photon time tagging shot noise expectation of the simulated rate and directly yields the RMSE correction factor of $0.31$ in this measurement. The RMSE correction factor is then multiplied to the Poissonian shot noise expectation to yield the current correlation RMSE expectation.

A detailed description of the general HPR setup and its measurement analysis chain can be found in \cite{10.1093/mnras/stab3058}

\section{Experiments and Results}
\label{sec:4}
The correlation results are analysed in two aspects. First, the measurement of the correlation time is analysed to investigate the coherence time. Second, the fluctuation of the correlation function baseline is analysed in the region around the peak to investigate possible background systematics in the measurement. A good index is the RMSE value of the $g^{(2)}$ baseline. Assuming no systematics, the RMSE value should be given by the Poissonian noise statistics.

\begin{figure*}
    \centering
    \includegraphics[width = \textwidth]{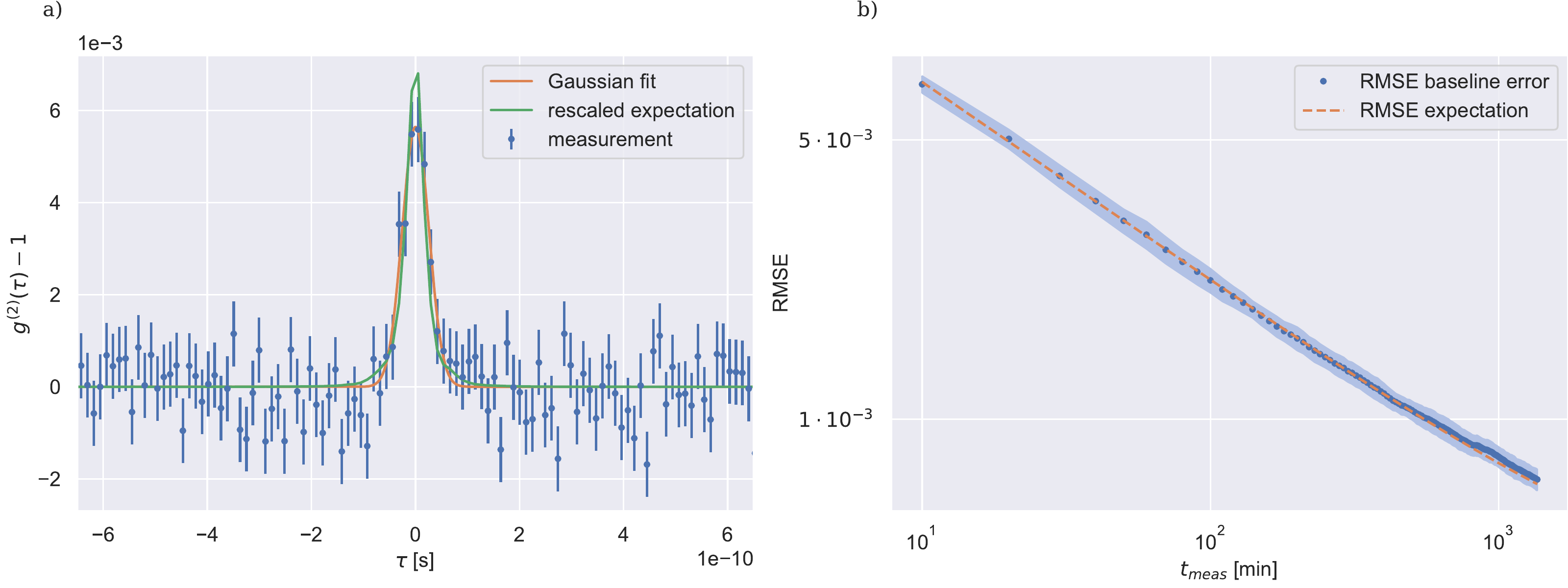}
    \caption{Measurement results for the HTR setup. Part a) shows the measured second order correlation function after an accumulation time of \SI{22.7}{\hour}, as well as a measurement expectation rescaled to the measured coherence time and a Gaussian fit. Part b) shows the HTR baseline RMSE time evolution over the accumulation time.}
    \label{fig:QOQI}
\end{figure*}

\begin{figure*}
    \centering
    \includegraphics[width = \textwidth]{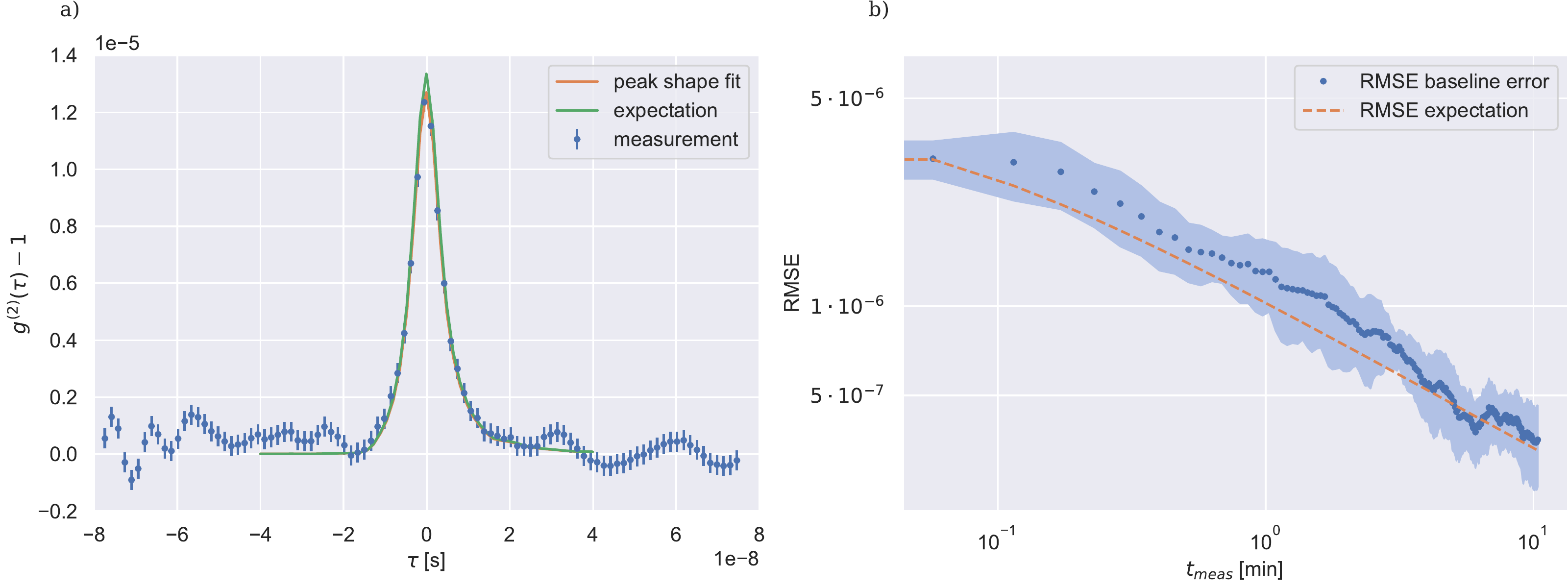}
    \caption{Measurement results for the HPR setup. Part a) shows the measured second order correlation function after an accumulation time of \SI{10}{\minute}, as well as the measurement expectation calculated numerically and a fit of that numerical model to the measured correlation function. Part b) shows the HPR baseline RMSE time evolution over the accumulation time.}
    \label{fig:ECAP}
\end{figure*}




Part a) of \cref{fig:QOQI} shows the correlation function measured by the HTR setup (\SI{0.5}{\nano \m} optical bandwidth, centred at \SI{416}{\nano \m}) after a measurement time of \SI{22.7}{\hour}, where the error bars correspond to the baseline RMSE.
Evaluating the coherence time by numerical integration we recover a value of about 85 percent of the expected coherence time ($\tau_\text{c, exp} = \SI{0.425}{\pico \s}$, $\tau_\text{c, meas} = \SI{0.35 +- 0.03}{\pico \s}$). The expected coherence time is calculated from a manufacturer-supplied filter transmission spectrum. A Gaussian fit to the measurement yields a coherence time of $\tau_\text{c, fit} = \SI{0.36 +- 0.03}{\pico \s}$, and is also shown in \cref{fig:QOQI}. The coherence times recovered by fit and numerical integration agree well with each other, leading to the conclusion that a coherence time about 15 percent lower than the expectation has been measured. Part a) of \Cref{fig:QOQI} also shows an expectation curve calculated from measurements of the HTR timing jitter and a Gaussian filter, rescaled to yield a coherence time matching the measured value. This curve is in good agreement with the measured correlation function, and only nearly exceeds the $1 \sigma$ boundary of the measurement values for values centred around zero time delay. The reduction in measured coherence time could be explained by slight errors in the manufacturer supplied filter transmission spectrum, which could fully explain the observed deviations even for a spectrometer FWHM resolution $< \SI{0.1}{\nano \m}$. 

Part b) of \cref{fig:QOQI} shows the RMSE behaviour of the HTR $g^{(2)}$ function evaluated cumulatively over the whole measurement time. The RMSE is evaluated directly on the measured correlation histogram baseline for each time period. In order to estimate the error of our calculated RMSE, it is evaluated in 20 independent samples of size more than 20 times larger than the measured correlation peak FWHM. The theoretical predictions for the expected RMSE is directly extracted from the mean number of counts per bin of both the raw and the normalization measurement, and are approximated very well by our measurements. 

Part a) of \cref{fig:ECAP} shows the correlation function obtained with the HPR setup using the XBO lamp filtered at \SI{465}{\nano \m} with a FWHM of \SI{2}{\nano \m}. The error bars correspond to the baseline RMSE.
Since timing resolution and sampling time are far larger than the correlation time, the expected $g^{(2)}$ signal shape is dominated by the correlation pulse shape of PMT pulses. 
The expected pulse shape is derived by correlating the average single photon pulse shapes of both channels obtained from a calibration measurement, as previously described. The resulting height of the correlation signal can then be scaled to fit the requirement of its integral being the coherence time as defined (labelled "expectation" in \cref{fig:ECAP}). Furthermore, the shape can also be fitted to the data with the height and the peak position being free fit parameters, and allows to derive the experimental value of the coherence time. Both are plotted in \cref{fig:ECAP} and show good agreement with each other. The theoretical coherence time stems from numerical calculations using the optical filter function measured with a high-resolution spectrometer.

Part b) of \cref{fig:ECAP} shows the RMSE of the analysed $g^{(2)}$ function evaluated cumulatively over the whole measurement time. The uncertainty is computed by taking $8$ independent parts of $g^{(2)}|_{\tau \gg \tau_\textnormal{c}}$ each of $25$ time bins length. 
The dashed line indicates the Poissonian shot noise statistics following nearly a $1/\sqrt{T}$ curve only slightly affected by the non-constant photon rates.
It is apparent that the measured RMSE values follow the theory curve quite well for the whole measurement time of \SI{10}{\min}. Other than shot noise, contributions are well below the $10^{-6}$ level.


\section{Discussion}
\label{sec:5}


\begin{figure}
    \centering
    \includegraphics[width=\columnwidth]{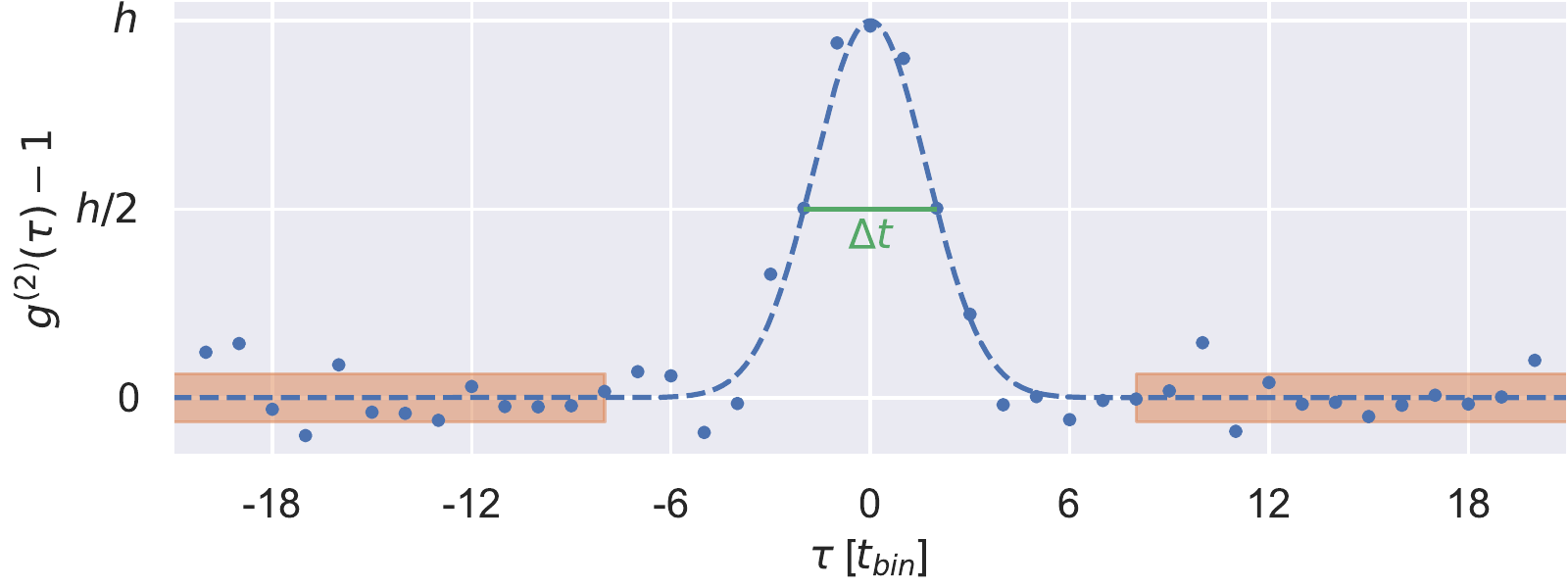}
    \caption{Illustration of parameter definitions for the S/N calculation. The bunching peak FWHM is shown in green and the RMSE interval around the second order correlation function baseline is shaded in orange.}
    \label{fig:sn_def}
\end{figure}

In this section, we compare the two methods with respect to S/N in a given measuring time.

It can be seen from the RMSE plots in \cref{fig:QOQI} and \cref{fig:ECAP} that both methods work at the minimum noise level, which means that they are dominated by the photon number Poisson noise (shot noise).
As presented in \cref{fig:QOQI} and \cref{fig:ECAP} for both methods we can determine the height $h$ of the $g^{(2)}$-correlation peak, the FWHM time width $\Delta t$ of the peak and the noise RMSE of the $g^{(2)}$-curve per time bin of bin width $t_{\text{bin}}$ 
The parameter definitions are also displayed in \cref{fig:sn_def}.

The signal-to-noise ratio of a correlation measurement can be approximated as
\begin{equation}
   \frac{S}{N} =  \frac{h}{\text{RMSE}}     \, \sqrt{\frac{\Delta t}{t_{\text{bin}}} },
\end{equation}
multiplying the peak height and RMSE error by the ratio of the correlation peak FWHM and correlation bin size to yield an effective integral of the correlation peak and its RMSE error.

As an example we compare the S/N values for the two setups at a measurement time of 10 minutes. The corresponding RMSE in the HTR (HPR) setup is the first (last) data point in \cref{fig:QOQI} (\cref{fig:ECAP}). The S/N ratios are

\begin{equation}
   \left(\frac{S}{N}\right)_\text{HTR} = \frac{0.0056}{0.0069} \, \sqrt{\frac{60\,\text{ps}}{12\,\text{ps}}}  = 1.8 
\end{equation}
\begin{equation}
      \left(\frac{S}{N}\right)_\text{HPR} = \frac{1.27\cdot10^{-5}}{3.51\cdot10^{-7}}    \, \sqrt{\frac{8\,\text{ns}}{1.6\,\text{ns}}}   = 80.9
\end{equation}

The difference between the two results of the different setups can be explained by the time resolution (resp. peak width) $\Delta t$ and the relative intensity $f_\text{trans}$ between the two setups. The relative intensity is determined by the applied grey filter $f_\text{trans, grey} = 0.028$, the difference in quantum efficiency $f_\text{trans, QE} = 0.5$ and the difference in source brightness at the measured wavelengths $f_\text{trans, spectrum} = 0.5$, all reducing the HTR photon flux relative to $f_\text{trans, HPR} = 1$.
Further, one has to correct for the different amount of temporal coherence intrinsically determined by the respective filter's central wavelength, $f_{\text{coh}}$.
The achieved S/N-values should be proportional to $f_\text{trans} f_{\text{coh}} /\sqrt{\Delta t}$.
For the HPR setup the effective time resolution differs from the peak width due to the net width of photon pulse correlations. Assuming no change in S/N between the concepts of time tagging and current correlation, $\Delta t_\text{HPR} = c^2 \cdot \Delta t_\text{peak}$ holds, where $c = 0.31$ is the previously mentioned RMSE correction factor.

With these values we expect
\begin{equation}
   \frac{\left(\frac{S}{N}\right)_\text{HTR}} {\left(\frac{S}{N}\right)_\text{HPR}} = \frac{(f_\text{trans} f_{\text{coh}} /\sqrt{\Delta t})_\text{HTR}}{(f_\text{trans} f_{\text{coh}} /\sqrt{\Delta t})_\text{HPR}} = 
   \frac{0.0075 \cdot 0.8 / \sqrt{60\,\text{ps}}}{1 \cdot 1 / \sqrt{0.76\,\text{ns}} } = 
   0.021
\end{equation}
which is in excellent agreement with the measured S/N values. \\

In our laboratory measurements, the HPR setup provides a much higher S/N in the same measurement time than the HTR setup. However, this is mainly because of the strong reduction of photon flux in the HTR setup, which is analogous to measuring the same star with a smaller telescope. From the rates measured in the laboratory, we extract the virtual telescope areas employed for our autocorrelation measurement. As an exemplary star we choose a magnitude 2 star and calculate that HPR is measuring with an equivalent of two $195$\,m$^2$ telescopes ($15.8\,$m diameter), while HTR uses two telescopes of size $1.17$\,m$^2$ ($1.2\,$m diameter). A total light collection and detection efficiency of $0.1$ is assumed here for both systems.

\begin{table}
\centering
\begin{tabular}{| c | c | c | c |} 
 \hline
 Telescope & Area (m$^2$) & $\Delta t$ (ns) & Reference \\ [0.5ex] 
 \hline\hline 
 H.E.S.S. I.   & 100 & 5 & \cite{bernlohr2003optical} \\
 CTA SST (req) &   8 & $1.5\times 2.35$ & \cite{rulten2016simulating}\\
 CTA MST (req) &  88 & $0.8\sqrt{12}$ & \cite{schlenstedt2014medium}\\
 CTA LST       & 370 & 0.1 & \\
 MAGIC         & 227 & 0.4 & \cite{shayduk2005new} \\
 VERITAS       & 113 &   4 & \cite{weekes2002veritas}\\
 \hline
 PW CDK 20   & 0.19 & $\approx 0.01$  & \cite{Cavazzani2012} \\
 PW CDK 1000 & 0.8  & $\approx  0.01$ & \cite{Cavazzani2012} \\
 \hline
 HTR (this work)    & 1.17 & 0.041  & \\
 HPR (this work)    & 202.77 & 0.76 & \\[1ex] 
 \hline
\end{tabular}
\caption{Summary of values for telescope areas and time resolutions used for \cref{fig:snr2D}. The CTA requirements are used for the CTA SST and MST (we assume a Gaussian distribution of photon arrival times for the SST and a rectangular distribution for the MST), for the CTA LST we refer to a private communication with D. Mazin. For the PW telescopes at least $\lambda / 4 $ precision is assumed due to their imaging capability, and the timing resolution of photons impinging on the detector is dominated by jitter introduced by atmospheric fluctuations \citep{Cavazzani2012}.}
\label{table:tel_data}
\end{table}

\begin{figure}
    \centering
    \includegraphics[width=\columnwidth]{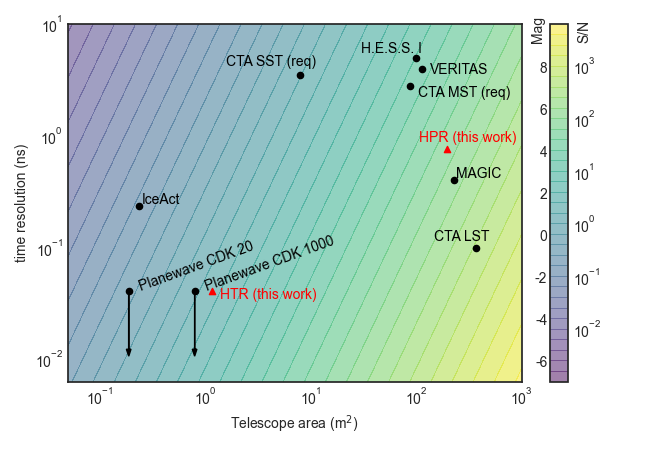}
    \caption{Map of telescope properties. Two quantities are shown on the colour bar. The right scale shows the $S/N$ as defined in this paper for a magnitude 2 star after $10\,$minutes of data accumulation using two telescopes of each type at $100\,$percent spatial coherence. 5 contour lines correspond to a factor of $10$ change. The red data points correspond to the measurements discussed above, plotted as hypothetical star measurements with telescope sizes matching the photon rates of the measurements. The left scale of the colour bar shows the limiting magnitude of a star requiring $S/N \geq 5$ for a $10\,$minutes measurement.
    Arrows pointing down denote an upper limit on the time resolution, set by the current HTR time resolution in combination with the Planewave telescopes.
    We assume a combined light collection and detection efficiency of $0.1$ for the S/N and magnitude map. This is a rough assumption not specific to any of the shown telescope arrays, thus there may be differences to (possible) realizations for each of the arrays.}
    \label{fig:snr2D}
\end{figure}

\Cref{table:tel_data} displays a few telescope types used in (upcoming) telescope arrays, two smaller optical telescopes as well as the the virtual telescopes of this work. The sizes and the telescope's optical time spreads are given. The data is visualized in \cref{fig:snr2D}.  

Telescopes suitable for HTR are located in the lower left corner of the plot providing small collection areas but high isochronicity and thus small optical time spread. Values for their isochronicity are assumed to be dominated by atmospheric turbulence as described in \cite{Cavazzani2012}, since the telescope itself displays imaging capabilities with unisochronicity $< \lambda/4$. The time resolution is thus dominated by the single photon detection equipment rather than the optical time spread of the telescope itself.
The HPR setup can be operated at very high photon rates, which doesn't intrinsically limit the collection area, however, usually such telescopes are parts of Cherenkov telescope arrays and have an unisochronicity of a few nanoseconds (even though with MAGIC and the CTA LST also high time-resolution Cherenkov telescopes exist/will be built). These telescopes are mainly located in the upper right corner in \cref{fig:snr2D}, their time resolutions are dominantly given by the telescope's unisochronicities. Due to the large point spread function (PSF) of Cherenkov telescopes, intensity interferometry at such telescopes would suffer from night sky background to a higher extent than for telescope types with a smaller PSF. The large night sky background contribution might set a limiting magnitude for observations much smaller than the limiting magnitudes estimated from our background-free laboratory measurements, especially for bright full-moon nights. Further cosmic ray produced Cherenkov light background could influence S/N for faint limiting magnitudes \citep{HanburyBrown1974}.

A S/N map for a measurement time of 10 minutes of a mag 2 star is displayed In \cref{fig:snr2D}, see the right numbers of the colour bar. For a more intuitive view when comparing the different telescopes the reader may also consider the left numbers on the colour bar, denoting the maximum magnitude of a star which satisfies $S/N \geq 5$ after $10$ minutes measurement time. However, please note that the red data-points do not carry meaning in this interpretation.

It can be seen  that with the HTR setup and two \SI{1}{\m} diameter telescopes, S/N is comparable to the ones achievable with two small Cherenkov telescopes such as the CTA SSTs. For mirrors on the order of \SI{3}{\m} diameter or $>7\,$m$^2$ of area, an HTR setup could compete with measurements taken at CTA MSTs. Such larger mirror diameters are feasible for use with single photon counting detectors for faint stars,by utilizing multiple single photon detectors and beam splitters per telescope to reduce the incident flux on each detector, or using detectors with dead times on the order of \SI{10}{\nano \s}. While such detectors with single photon resolution are currently under development \citep{Orlov2019a, Orlov2019} they are not yet widely available. Another promising detector type for intensity interferometry are superconducting nanowire single photon detectors (SNSPDs) due to their high quantum efficiency, low dead time, and high timing resolution, for which efficient coupling to multimode fibers has been demonstrated recently (see e.g. \cite{Chang2019}). While SNSPDs are commercially avauliable, they would require effcient fiber coupling at the telescope, requiring high-grade optics. This means, that currently the raw S/N favours the current-correlating HPR approach for IACTs. However, an array of $1\,$m telescopes used in conjunction with a HTR-like setup could alternatively be used fruitfully in the future.

Assuming the time resolution of CTA-MSTs in combination with HPR intensity interferometry electronics being not better than the MST's requirement (an admittedly conservative estimation), we compare use of 2 MSTs with HPR electronics to an array of optical \SI{1}{\m} diameter telescopes. For this comparison, the availability of different telescope baselines when using an array of telescopes is not subject of investigation, instead the individual telescopes add up to a combined light collection area. We calculate an S/N value of $36.5$ for the HPR-MSTs, which corresponds to a measurement with 26 pairs of \SI{1}{\m} telescopes (S/N = $36.0$) and thus only 23 percent of the mirror area of two MSTs, demonstrating the potential of high time resolution equipment for intensity correlation measurements. Employing an HTR-like HBT instrumentation with higher quantum efficiency as for example found in Si APDs, the number of \SI{1}{m} telescopes necessary to match the two MST S/N  could be reduced by a factor of two. As suggested in \cite{Kim2014}, the sensitivity of both intensity interferometry setups could be enhanced by spectral and polarization multiplexing, leading to accessible apparent magnitudes below 10. This enhancement would probably scale especially favourably for a HTR-like approach fielding narrow spectral filtering and a high quality beam that can be easily controlled.


Note also that the small telescopes could be arranged in flexible geometries, and might thus prove advantageous for the extraction of specific features of astronomical objects. Further, they could realize larger baselines and thus potentially obtain a higher angular resolution than CTA telescopes, again since they are not constrained to the geometry of CTA. In particular, they could be used to supplement the results gained from intensity interferometry with CTA telescopes.

Given that CTA-S will be built, also the use of all 14 planned MSTs can be considered. The HPR-S/N then corresponds to $255$ which would require $368$ HTR telescopes to achieve a similar S/N. Not even taking LSTs and SSTs with the additive longer baselines into this consideration it demonstrates that with good reason CTA-S is considered to be a most-promising concept for successful intensity interferometry operations coming up in the future decade(s).


\section{Outlook}

With the presented laboratory measurements, we have proven that both portrayed methods can work at the photon noise limit. The photon counting (HTR) ansatz aims at highest time resolution with moderate photon rates. It achieved \SI{60}{\pico \s} time resolution maintained over several hours. The related noise level is $10^{-3}$. The photon current (HPR) ansatz aims at highest PMT current and moderate time resolution. It achieved a rate of \SI{950}{\mega \hertz} and a stable noise level of $10^{-6}$ over several minutes.\\
Given that both methods work at the ideal noise limits, they deliver predictable signal-to-noise levels for astronomical observations with reasonable telescope fields. In particular, it has been shown that small HTR telescopes can compete with HPR at bigger IACTs. Based on the availability of (future) IACT arrays HPR-implementations seem to be encouraging science cases. However, further developments of HTR electronics to higher rate-capabilities promise to stay competitive.

\section*{Acknowledgements}

This work was supported with a grant by the Deutsche Forschungsgemeinschaft (\lq Optical intensity interferometry with the H.E.S.S. gamma-ray telescopes\rq - FU 1093/3-1, Project number 426212122). The authors thank Heide Constantini and Nolan Matthews for their helpful comments on the manuscript. 


\section*{Data Availability}
The data that support the plots within this paper and other findings of this study
are available from the corresponding authors upon reasonable request. For the HTR setup all intermediary correlation histograms are available in time-intervals of \SI{30}{\s}, however no raw photon event data can be supplied. For the HPR all intermediary correlation histograms are available in time-intervals of \SI{3.436}{\s}. Due to the large size of the digitized waveforms in excess of a TB, the raw data cannot be made available online.



\bibliographystyle{mnras}
\bibliography{example} 

\begin{thebibliography}{}
\makeatletter
\relax
\def\mn@urlcharsother{\let\do\@makeother \do\$\do\&\do\#\do\^\do\_\do\%\do\~}
\def\mn@doi{\begingroup\mn@urlcharsother \@ifnextchar [ {\mn@doi@}
  {\mn@doi@[]}}
\def\mn@doi@[#1]#2{\def\@tempa{#1}\ifx\@tempa\@empty \href
  {http://dx.doi.org/#2} {doi:#2}\else \href {http://dx.doi.org/#2} {#1}\fi
  \endgroup}
\def\mn@eprint#1#2{\mn@eprint@#1:#2::\@nil}
\def\mn@eprint@arXiv#1{\href {http://arxiv.org/abs/#1} {{\tt arXiv:#1}}}
\def\mn@eprint@dblp#1{\href {http://dblp.uni-trier.de/rec/bibtex/#1.xml}
  {dblp:#1}}
\def\mn@eprint@#1:#2:#3:#4\@nil{\def\@tempa {#1}\def\@tempb {#2}\def\@tempc
  {#3}\ifx \@tempc \@empty \let \@tempc \@tempb \let \@tempb \@tempa \fi \ifx
  \@tempb \@empty \def\@tempb {arXiv}\fi \@ifundefined
  {mn@eprint@\@tempb}{\@tempb:\@tempc}{\expandafter \expandafter \csname
  mn@eprint@\@tempb\endcsname \expandafter{\@tempc}}}

\bibitem[\protect\citeauthoryear{Abeysekara et~al.,}{Abeysekara
  et~al.}{2020}]{veritas_II}
Abeysekara A.~U.,  et~al., 2020, \mn@doi [Nature Astronomy]
  {10.1038/s41550-020-1143-y}

\bibitem[\protect\citeauthoryear{Acciari et~al.,}{Acciari
  et~al.}{2019}]{magic_ii}
Acciari V.~A.,  et~al., 2019, \mn@doi [Monthly Notices of the Royal
  Astronomical Society] {10.1093/mnras/stz3171}, 491, 1540

\bibitem[\protect\citeauthoryear{Baines, van Belle, ten Brummelaar, McAlister,
  Swain, Turner, Sturmann  \& Sturmann}{Baines et~al.}{2007}]{Baines2007}
Baines E.~K.,  van Belle G.~T.,  ten Brummelaar T.~A.,  McAlister H.~A.,  Swain
  M.,  Turner N.~H.,  Sturmann L.,   Sturmann J.,  2007, \mn@doi [The
  Astrophysical Journal] {10.1086/519002}, 661, L195

\bibitem[\protect\citeauthoryear{{Becker \& Hickl GmbH}}{{Becker \& Hickl
  GmbH}}{2017}]{BHG2017}
{Becker \& Hickl GmbH} 2017, {HPM-100-06/07: Ultra-High Speed Hybrid Detectors
  for TCSPC}, \url
  {https://www.becker-hickl.com/wp-content/uploads/2018/11/db-hpm-06-07-v02.pdf}

\bibitem[\protect\citeauthoryear{Bernl{\"o}hr et~al.,}{Bernl{\"o}hr
  et~al.}{2003}]{bernlohr2003optical}
Bernl{\"o}hr K.,  et~al., 2003, Astroparticle Physics, 20, 111

\bibitem[\protect\citeauthoryear{Bohec \& Holder}{Bohec \&
  Holder}{2006}]{Bohec2006}
Bohec S.~L.,  Holder J.,  2006, \mn@doi [The Astrophysical Journal]
  {10.1086/506379}, 649, 399

\bibitem[\protect\citeauthoryear{Brown \& Twiss}{Brown \&
  Twiss}{1956}]{BROWN1956}
Brown R.~H.,  Twiss R.~Q.,  1956, \mn@doi [Nature] {10.1038/177027a0}, 177, 27

\bibitem[\protect\citeauthoryear{Brown, Twiss, Lovell  \& surName}{Brown
  et~al.}{1957}]{Brown1957}
Brown R.~H.,  Twiss R.~Q.,  Lovell A. C.~B.,   surName g.,  1957, \mn@doi
  [Proceedings of the Royal Society of London. Series A. Mathematical and
  Physical Sciences] {10.1098/rspa.1957.0177}, 242, 300

\bibitem[\protect\citeauthoryear{Brown, Davis  \& Allen}{Brown
  et~al.}{1967}]{brown1967stellar}
Brown R.~H.,  Davis J.,   Allen L.,  1967, Monthly Notices of the Royal
  Astronomical Society, 137, 375

\bibitem[\protect\citeauthoryear{Cavazzani, Ortolani  \& Barbieri}{Cavazzani
  et~al.}{2012}]{Cavazzani2012}
Cavazzani S.,  Ortolani S.,   Barbieri C.,  2012, \mn@doi [Monthly Notices of
  the Royal Astronomical Society] {10.1111/j.1365-2966.2011.19883.x}, 419, 2349

\bibitem[\protect\citeauthoryear{Chang et~al.,}{Chang et~al.}{2019}]{Chang2019}
Chang J.,  et~al., 2019, \mn@doi [Appl. Opt.] {10.1364/AO.58.009803}, 58, 9803

\bibitem[\protect\citeauthoryear{Che et~al.,}{Che et~al.}{2011}]{Che2011}
Che X.,  et~al., 2011, \mn@doi [The Astrophysical Journal]
  {10.1088/0004-637x/732/2/68}, 732, 68

\bibitem[\protect\citeauthoryear{Dravins, LeBohec, Jensen  \& Nuñez}{Dravins
  et~al.}{2013}]{Dravins2013}
Dravins D.,  LeBohec S.,  Jensen H.,   Nuñez P.~D.,  2013, \mn@doi
  [Astroparticle Physics]
  {https://doi.org/10.1016/j.astropartphys.2012.04.017}, 43, 331

\bibitem[\protect\citeauthoryear{Fox}{Fox}{2006}]{Fox2006}
Fox M.,  2006, Quantum optics: an introduction.
Oxford Master Series in Physics, 6, Oxford University Press, USA, \url
  {http://gen.lib.rus.ec/book/index.php?md5=AF074928C421227E6B7EBFCC30425DE2}

\bibitem[\protect\citeauthoryear{Guerin, Dussaux, Fouch{\'e}, Labeyrie, Rivet,
  Vernet, Vakili  \& Kaiser}{Guerin et~al.}{2017}]{guerin2017temporal}
Guerin W.,  Dussaux A.,  Fouch{\'e} M.,  Labeyrie G.,  Rivet J.-P.,  Vernet D.,
   Vakili F.,   Kaiser R.,  2017, Monthly Notices of the Royal Astronomical
  Society, 472, 4126

\bibitem[\protect\citeauthoryear{Guerin, Rivet, Fouch{\'e}, Labeyrie, Vernet,
  Vakili  \& Kaiser}{Guerin et~al.}{2018}]{guerin2018spatial}
Guerin W.,  Rivet J.-P.,  Fouch{\'e} M.,  Labeyrie G.,  Vernet D.,  Vakili F.,
   Kaiser R.,  2018, Monthly Notices of the Royal Astronomical Society, 480,
  245

\bibitem[\protect\citeauthoryear{Hanbury~Brown}{Hanbury~Brown}{1974}]{HanburyBrown1974}
Hanbury~Brown R.,  1974, The intensity interferometer; its application to
  astronomy.
Taylor \& Francis, Halsted Press, \url
  {http://gen.lib.rus.ec/book/index.php?md5=d604cdb5cacf1ea40203b4b71256ea19}

\bibitem[\protect\citeauthoryear{Hanbury~Brown, Davis  \& Allen}{Hanbury~Brown
  et~al.}{1974}]{HBT_32}
Hanbury~Brown R.,  Davis J.,   Allen L.~R.,  1974, \mn@doi [Monthly Notices of
  the Royal Astronomical Society] {10.1093/mnras/167.1.121}, 167, 121

\bibitem[\protect\citeauthoryear{Horch, Weiss, Klaucke, Pellegrino  \&
  Rupert}{Horch et~al.}{2021}]{Horch2021}
Horch E.~P.,  Weiss S.~A.,  Klaucke P.~M.,  Pellegrino R.~A.,   Rupert J.~D.,
  2021, Observations with the Southern Connecticut Stellar Interferometer. I.
  Instrument Description and First Results (\mn@eprint {arXiv} {2112.07758})

\bibitem[\protect\citeauthoryear{Labeyrie, Lipson  \& Nisenson}{Labeyrie
  et~al.}{2006}]{Labeyrie2006}
Labeyrie A.,  Lipson S.~G.,   Nisenson P.,  2006, An Introduction to Optical
  Stellar Interferometry.
Cambridge University Press, Cambridge, \mn@doi{10.1017/CBO9780511617638}, \url
  {https://www.cambridge.org/core/books/an-introduction-to-optical-stellar-interferometry/2EA3ABDA8557CF3277063391C02E899D}

\bibitem[\protect\citeauthoryear{Loudon}{Loudon}{2000}]{Loudon2000}
Loudon R.,  2000, The Quantum Theory of Light, 3rd ed. (Oxford Science
  Publications), 3 edn.
Oxford University Press, USA, \url
  {libgen.li/file.php?md5=ccc4545c9d7c0a3216fb446644bfcd47}

\bibitem[\protect\citeauthoryear{Maier}{Maier}{2007}]{maier2007veritas}
Maier G.,  2007, VERITAS: Status and Latest Results (\mn@eprint {arXiv}
  {0709.3654})

\bibitem[\protect\citeauthoryear{Mandel \& Wolf}{Mandel \&
  Wolf}{1995}]{Mandel1995}
Mandel L.,  Wolf E.,  1995, Optical Coherence and Quantum Optics.
Cambridge University Press, \mn@doi{10.1017/CBO9781139644105}

\bibitem[\protect\citeauthoryear{{Mourard, D.} et~al.,}{{Mourard, D.}
  et~al.}{2015}]{Mourard2015}
{Mourard, D.} et~al., 2015, \mn@doi [A\&A] {10.1051/0004-6361/201425141}, 577,
  A51

\bibitem[\protect\citeauthoryear{Nuñez \& Domiciano~de Souza}{Nuñez \&
  Domiciano~de Souza}{2015}]{Nunez2015}
Nuñez P.~D.,  Domiciano~de Souza A.,  2015, \mn@doi [Monthly Notices of the
  Royal Astronomical Society] {10.1093/mnras/stv1719}, 453, 1999

\bibitem[\protect\citeauthoryear{Nuñez, Holmes, Kieda, Rou  \& LeBohec}{Nuñez
  et~al.}{2012}]{Nunez2012}
Nuñez P.~D.,  Holmes R.,  Kieda D.,  Rou J.,   LeBohec S.,  2012, \mn@doi
  [Monthly Notices of the Royal Astronomical Society]
  {10.1111/j.1365-2966.2012.21263.x}, 424, 1006

\bibitem[\protect\citeauthoryear{Orlov, Glazenborg, Ortega  \& Kernen}{Orlov
  et~al.}{2019a}]{Orlov2019a}
Orlov D.~A.,  Glazenborg R.,  Ortega R.,   Kernen E.,  2019a, \mn@doi [CEAS
  Space Journal] {10.1007/s12567-019-00260-0}, 11, 405

\bibitem[\protect\citeauthoryear{Orlov, Glazenborg, Ortega  \& Kernen}{Orlov
  et~al.}{2019b}]{Orlov2019}
Orlov D.~A.,  Glazenborg R.,  Ortega R.,   Kernen E.,  2019b, in Itzler M.~A.,
  Bienfang J.~C.,   McIntosh K.~A.,  eds, ~ Vol. 10978, Advanced Photon
  Counting Techniques XIII. SPIE, pp 93 -- 100, \url
  {https://doi.org/10.1117/12.2519061}

\bibitem[\protect\citeauthoryear{Rivet et~al.,}{Rivet et~al.}{2020}]{Rivet2020}
Rivet J.-P.,  et~al., 2020, \mn@doi [Monthly Notices of the Royal Astronomical
  Society] {10.1093/mnras/staa588}, 494, 218

\bibitem[\protect\citeauthoryear{Roettenbacher et~al.,}{Roettenbacher
  et~al.}{2016}]{Roettenbacher2016}
Roettenbacher R.~M.,  et~al., 2016, \mn@doi [Nature] {10.1038/nature17444},
  533, 217

\bibitem[\protect\citeauthoryear{Rulten, Zech, Okumura, Laporte  \&
  Schmoll}{Rulten et~al.}{2016}]{rulten2016simulating}
Rulten C.,  Zech A.,  Okumura A.,  Laporte P.,   Schmoll J.,  2016,
  Astroparticle Physics, 82, 36

\bibitem[\protect\citeauthoryear{Schlenstedt}{Schlenstedt}{2014}]{schlenstedt2014medium}
Schlenstedt S.,  2014, in Ground-based and Airborne Telescopes V. p. 91450O

\bibitem[\protect\citeauthoryear{Shayduk, Hengstebeck, Kalekin, Pavel  \&
  Schweizer}{Shayduk et~al.}{2005}]{shayduk2005new}
Shayduk M.,  Hengstebeck T.,  Kalekin O.,  Pavel N.,   Schweizer T.,  2005, in
  29th International Cosmic Ray Conference (ICRC29), Volume 5. p.~223

\bibitem[\protect\citeauthoryear{Tan, Yeo, Poh, Chan  \& Kurtsiefer}{Tan
  et~al.}{2014}]{Tan2014}
Tan P.~K.,  Yeo G.~H.,  Poh H.~S.,  Chan A.~H.,   Kurtsiefer C.,  2014, \mn@doi
  [The Astrophysical Journal] {10.1088/2041-8205/789/1/l10}, 789, L10

\bibitem[\protect\citeauthoryear{{Thorlabs, Inc}}{{Thorlabs,
  Inc}}{2021}]{ThorlabsNDFilter}
{Thorlabs, Inc} accessed: 06/2021, {Unmounted Absorptive Neutral Density
  Filters - Transmission and Reflection Graphs}, \url
  {https://www.thorlabs.com/newgrouppage9.cfm?objectgroup_id=5011}

\bibitem[\protect\citeauthoryear{Trippe et~al.,}{Trippe et~al.}{2014}]{Kim2014}
Trippe S.,  et~al., 2014, \mn@doi [Journal of The Korean Astronomical Society]
  {10.5303/JKAS.2014.47.6.235}, 47, 235

\bibitem[\protect\citeauthoryear{Weekes et~al.,}{Weekes
  et~al.}{2002}]{weekes2002veritas}
Weekes T.,  et~al., 2002, Astroparticle Physics, 17, 221

\bibitem[\protect\citeauthoryear{Zampieri et~al.,}{Zampieri
  et~al.}{2016}]{aqueye_ii}
Zampieri L.,  et~al., 2016, in Malbet F.,  Creech-Eakman M.~J.,   Tuthill
  P.~G.,  eds, ~ Vol. 9907, Optical and Infrared Interferometry and Imaging V.
  SPIE, pp 140 -- 153, \mn@doi{10.1117/12.2233688}, \url
  {https://doi.org/10.1117/12.2233688}

\bibitem[\protect\citeauthoryear{Zmija, Vogel, Anton, Malyshev, Michel, Zink
  \& Funk}{Zmija et~al.}{2021}]{10.1093/mnras/stab3058}
Zmija A.,  Vogel N.,  Anton G.,  Malyshev D.,  Michel T.,  Zink A.,   Funk S.,
  2021, \mn@doi [Monthly Notices of the Royal Astronomical Society]
  {10.1093/mnras/stab3058}, 509, 3113

\bibitem[\protect\citeauthoryear{ten Brummelaar et~al.,}{ten Brummelaar
  et~al.}{2005}]{CHARA_description}
ten Brummelaar T.~A.,  et~al., 2005, \mn@doi [The Astrophysical Journal]
  {10.1086/430729}, 628, 453–465

\makeatother
\end{thebibliography}








\bsp	
\label{lastpage}
\end{document}